\documentclass[aps,superscriptaddress,showpacs]{revtex4}
\usepackage{amssymb,epsfig,latexsym}
\usepackage{amsmath,amscd}
\usepackage{graphicx}
\usepackage[normalem]{ulem}
\usepackage[dvips]{color}

\begin{document}

\title{One possible mechanism for massive neutron star supported by  soft EOS}

\author{De-Hua Wen\footnote{email:wendehua@scut.edu.cn},  Jing Yan and Xue-Mei Liu}
 \affiliation{ Department of Physics, South China University of
Technology, Guangzhou  510641, P. R. China }


\begin{abstract}
The recently discovery of a massive neutron star (PSR J1614-2230
of $1.97\pm0.04M_{\odot}$) rules out the soft equation of states
(EOSs) such as those included hyperons or kaon condensates at high
densities, while the nuclear theory or the terrestrial laboratory
data prefer a soft EOS. Here we propose one possible mechanism to
allow that the observed massive neutron star can be supported by a
soft EOS, that is, if the the gravitational constant $G$ varies at
super strong field, a soft EOS can support the massive neutron
stars.
\end{abstract}

 \maketitle

\section{Introduction}
As early as in 1937, Dirac pointed out that the gravitational
``constant" $G$ should not be a universal constant and it will
decrease with the time \cite{Dirac}. After that, there are
numerous literatures are involved in the alterable gravitational
constant, see \cite{Uzan03,Peebles03,
Moss10,Singh09,Singh11,Bali11} etc. and references therein. Most
of them focused on the time dependence of the gravitational
constant, constrained by the different system, it is concluded
that the time dependence of the gravitational constant is around
$\dot{G}/G < 10^{-11}$ yr$^{-1}$ \cite{Uzan03}. While also some
authors argue that the gravitational constant should be replaced
by a scalar field which can vary both in space and time, say, the
gravitational constant in the Jordan-Fierz-Brans-Dicke theory
\cite{Brans61,Uzan03}. One recent work  shows that by analyzing
with a specifically selected equation of state (EOS), the
discovery of a two-solar mass neutron star provided a constraint
on the Newtonian gravitational constant, that is,  it cannot
exceed its value on Earth by more than 0.08 in the neutron star
\cite{Dobado11}.

Through a roughly qualitative estimation, the acceleration of
gravity $g$ on the surface of the neutron star is about $10^{12}~
m/s^{2}$, which is far larger than that of the earth, $\sim 9.8
~m/s^{2}$. In such a totally  different gravity circumstance, the
possibility of a variational  gravitational ``constant" could be
existent. On the other hand, the recently discovery of a massive
neutron star (PSR J1614-2230 with $1.97\pm0.04M_{\odot}$) rules
out the soft EOSs  \cite{Demorest10}, while both the nuclear
theory and the terrestrial laboratory data prefer softer EOSs
\cite{s15,s16,Xiao09,xu10}, where
 hyperons or kaon condensates may appear at high densities.

 Stimulated by the
very recent work \cite{Dobado11} in which a constraint on the
gravitational constant at strong field was put forward through the
observation of  massive neutron star, here we propose one possible
mechanism to allow the observed massive neutron star to be
supported by the soft EOS, say, an EOS for the neutron star
matters including hyperons described by the relativistic
$\sigma-\omega-\rho$ model (see the details in Section II). It is
worth noting that the EOSs of the dense matters are still
significantly model dependent up to date, for example,  recent
works show that even if hyperons exist in the stellar core, it is
still allowed the neutron star has a maximum stellar mass larger
than about $2.0~M_{\odot}$ \cite{Bednarek, Bonanno}. In fact, on
the one hand, scientists manage to let the EOS of the dense matter
become stiffer to resolve the puzzle that a soft EOS can not
support the observed massive neutron stars \cite{Bednarek,
Bonanno}. On the onther hand, there are also lots of efforts try
to find a reasonable mechanism to let a soft EOS of the dense
matter support the observed massive neutron star. One of the
successful methods is considering the non-Newtonian gravity in the
soft EOSs, which is equivalent to add a repulsive interaction in
the dense matters and thus to let the EOS become stiffer
\cite{wen09,wen11}.

\section{The equation of state described by relativistic $\sigma-\omega-\rho$ model}
\begin{figure}
\includegraphics{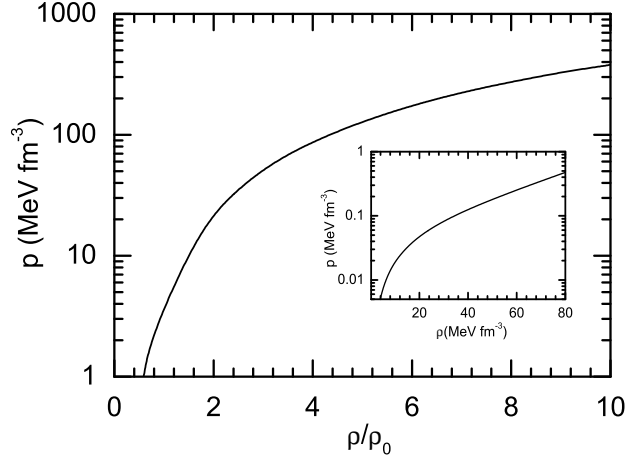}
\caption{\label{EOS} The equation of state of the neutron star
matters including hyperons, where the energy density $\rho$ is in
units of the saturation density $\rho_{0}$, and the insert shows
the EOS of the crust. }
\end{figure}

 It is widely believed that at
a density up to a few times the nuclear saturation density, the
exotic hadronic matter such as hyperons or kaon condensates will
appear, which  makes the EOS of the neutron star matters become
significant softer than those only include neutrons, protons and
electrons (npe)\cite{s15,s16,Burgio2007,Li2007}. As an example,
here we employ one soft EOS   investigated by the relativistic
$\sigma-\omega-\rho$ model, where hyperons are included at high
densities.
 This model is described by a Lagrangian
density as \cite{Johnson55,Duerr56,Walecka74,Schaffner96,
Glendenning01,Wen10}

\begin{eqnarray}\label{Lagrangian}
{
\textit{L}}&=&\sum_{B}\bar{\psi}_{B}(i\gamma_{\mu}\partial^{\mu}+m_{B}-g_{\sigma
B}\sigma-
g_{\omega B}\gamma_{\mu}\omega^{\mu}-\frac{1}{2}g_{\rho B}\gamma_{\mu}\vec{\tau}.\vec{\rho}^{\mu})\psi_{B}+\frac{1}{2}(\partial\sigma)^{2}\nonumber \\
& &-\frac{1}{2}m_{\sigma}^{2}\sigma^{2} -
\frac{1}{4}F_{\mu\nu}F^{\mu\nu}+\frac{1}{2}m_{\omega}^{2}\omega_{\mu}\omega^{\mu}-U(\sigma)-\frac{1}{4}\vec{\rho}_{\mu\nu}.\vec{\rho}^{\mu\nu}+
\frac{1}{2}m_{\rho}^{2}\vec{\rho}_{\mu}.\vec{\rho}^{\mu}\nonumber
\\
& &+
 \sum_{l}\bar{\psi}_{l}(i\gamma_{\mu}\partial^{\mu}-m_{l})\psi_{l}
\end{eqnarray}
where
\begin{equation}
U(\sigma)=a\sigma+\frac{1}{3!}c\sigma^{3}+\frac{1}{4!}d\sigma^{4},
\end{equation}

\begin{equation}
F_{\mu\nu}=\partial_{\mu}\omega_{\nu}-\partial_{\nu}\omega_{\mu},
\end{equation}

\begin{equation}
\vec{\rho}_{\mu\nu}=\partial_{\mu}\vec{\rho}_{\nu}-\partial_{\nu}\vec{\rho}_{\mu},
\end{equation}
$\psi_{B}$ is the field operator of baryon $B$ ($B$=$n$, $p$,
$\Lambda$, $\Sigma$, $\Xi$, $\Delta$); $\psi_{l}$ is the field
operator of lepton $l$ ($l$=$e$, $\mu$);   $\sigma$,
$\omega^{\mu}$, $\vec{\rho}^{\mu}$ are the field operators of
$\sigma$-, $\omega$-, $\rho$- meson, respectively; $g_{\sigma B}$,
$g_{\omega B}$, $g_{\rho B}$ are the coupling constants between
$\sigma$-, $\omega$-, $\rho$- meson and baryon, respectively;
$m_{B},m_{l},m_{i},(i=\sigma,\omega,\rho)$ are the mass of baryon,
lepton, meson, respectively; and $\vec{\tau}$ is the isospin
operator.  From the Lagrangian density described by Eq.
\ref{Lagrangian},  the EOS of the neutron star matters can be
obtained. In the numerical calculation, the following parameters
are adopted
  \cite{Wen10}: $a=-2.1\times 10^{7}~MeV^{3}$,
$c=0.97 M_{n}$, $d=1277$, $g_{s}=6.73$, $g_{v}=8.59$,
$M_{n}=938~MeV$, $m_{\omega}=783~MeV$, $m_{\sigma}=550~MeV$,
$m_{\rho}=770~MeV$ and $K=224~MeV$. Similar to other models, the
parameters adopted here are obtained by fitting and reproducing
the saturated properties and the symmetric energy of the nuclear
matter. The EOS  is displayed in Fig. \ref{EOS}. For the density
below about $0.07~fm^{-3}$, the results of Refs. \cite{s20,s21}
are employed, which is shown in the inset of Fig. 1.

\section{ Numerical results and discussion}

Before presenting the numerical results, we  first briefly review
the structure equation of the static neutron stars.
 Here we only consider the
relativistic hydrostatic equilibrium cases. The equilibrium of a
spherical perfect fluid star is described by a static, spherically
symmetric space-time with metric of the form
\begin{equation}
ds^{2}=-e^{2\nu}dt^{2}+e^{2\lambda}dr^{2}+r^{2}(d\theta^{2}+\sin^{2}\theta
d\phi^{2}),
\end{equation}
where $\nu$ and $\lambda$ are the functions of $r$ only.

\indent Supposing the matter of a static  spherically symmetric
neutron star can be treated as perfect fluid, therefore its
energy-momentum tensor can be described by
\begin{equation}
T_{\mu\nu}=pg_{\mu\nu}+(p+\rho)u_{\mu}u_{\nu},
\end{equation}
then according to the Einstein field equation
\begin{equation}
R_{\mu\nu}-\frac{1}{2}Rg_{\mu\nu}=-8\pi GT_{_{\mu\nu}},
\end{equation}
the Tolman-Oppenheimer-Volkoff (TOV) equations of the relativistic
hydrostatic equilibrium can be obtained \cite{Tolman, Oppenheimer}
\begin{equation}\label{eq8}
 \frac{dp}{dr}=-\frac{G}{c^{2}}\frac{(p+\rho)[m(r)+4\pi
 r^{3}p]}{r[r-2Gm(r)/c^{2}]}.
\end{equation}
Here the gravitational constant $G$ in Eq. \ref{eq8} is considered
as a  alterable ``constant". In the numerical calculation for the
neutron star structure,  the variational $G$ is denoted by the
ratio $G/G_{0}$, where $G_{0} (=6.6738\times 10^{-11}m^{3}\cdot
kg^{-1}\cdot s^{-2})$ is the Newtonian gravitational constant on
earth.

\begin{figure}
\includegraphics{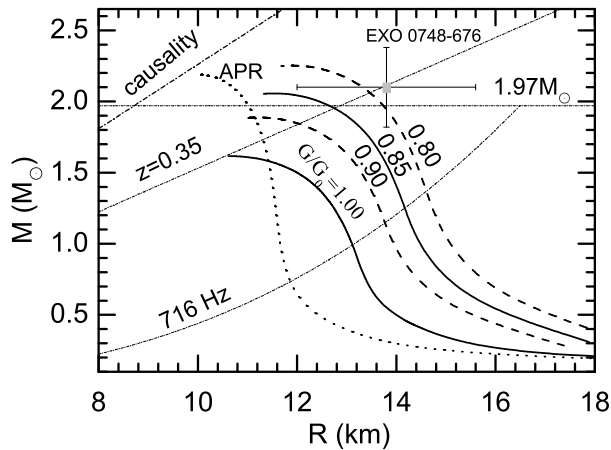}
 \caption{\label{TOV}   The
mass-radius relation of static neutron stars with different
gravitational constant $G$. The value $G/G_{0}$ is indicted by the
numbers beside the lines, where $G_{0}=6.6738\times
10^{-11}(m^{3}kg^{-1} s^{-2}$).}
\end{figure}

Shown in Fig.2 is the mass-radius relation of static neutron stars
with varying gravitational constant obtained from solving the TOV
equations. it is shown that if we do not consider the variation of
the gravitational constant ( $G/G_{0}=1$), the corresponding
sequence neutron star  can only support a maximum mass about $1.61
M_{\odot}$, which is obviously can not support the observed mass
$1.97\pm0.04 M_{\odot}$ of  PSR J1614-2230. In addition, as shown
in Fig. \ref{TOV}, for this sequence of neutron star, it is hard
to support a redshift $z=0.35$ \cite{Cottam02}. In order to
support a stellar mass up to $1.97 M_{\odot}$, one needs the
gravitational constant decreasing down to about $G/G_{0}=0.87$. If
we want the neutron star sequence can support the observation of
EXO 0748-676 (with $M\geq 2.10\pm0.28 M_{\odot}$ and $R\geq
13.8\pm 1.8km$) \cite{Ozel06}, the gravitational constant should
best decline to about $G/G_{0}=0.80$. For comparison, the
mass-radius relation of the widely used APR EOS \cite{s22}, which
consists of neutrons, protons, electrons and muons, is also
plotted in Fig.2, where the variation of the gravitational
constant is not considered. One can see that though the  neutron
star sequence of APR EOS can support the mass observation of PSR
J1614-2230, it can not support the observed radius of EXO
0748-676.

\vspace*{4mm}
\begin{figure}
\centerline{\includegraphics{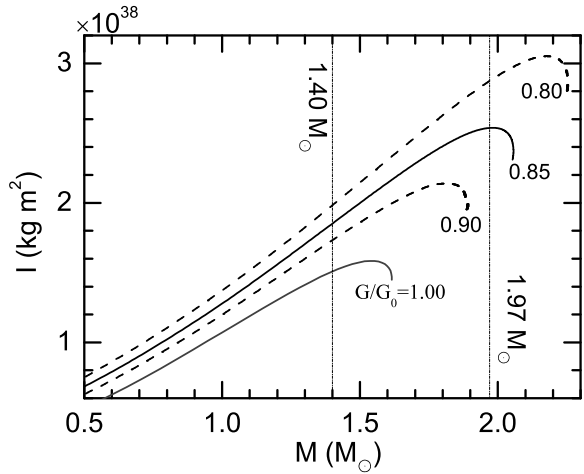}}
\begin{center}
\parbox{15.5cm}{\small  \caption{\label{M-I}  The momenta of inertia of the slowly
rotating hyperon stars with different gravitational constant
(denoting by the numbers of $G/G_{0}$ below the endpoints of
lines) as a function of the stellar masses. } }
\end{center}
\end{figure}

As the moment of inertia  can provide an important observational
constraints for the neutron stars, and the discovery of the
double-pulsar system PSR J0737-3039 $A\&B$ provides a great
opportunity to determine accurately the moment of inertia $I$ of
the star $A$ \cite{Lyne04, Lattimer07}, so it is also interesting
to investigate the effect of the gravitational ``constant" $G$ on
the moment of inertia.
 The moment of inertia is defined by
 \begin{equation}
I=\frac{J}{\Omega}
\end{equation}
where  $\Omega$ is the star's angular velocity, and $J$ is the
angular momentum. For the rotational frequency much lower than the
Kepler frequency, approximating to the first order terms in
 $\Omega$ , the moment of inertia can be estimated by an
available empirical equation \cite{s28}
  \begin{equation}
I\approx0.237~MR^{2}[1+4.2\frac{M~km}{M_{\odot}~R}+90(\frac{M~km}{M_{\odot}~R})^{4}].
\end{equation}

The momenta of inertia of the neutron star sequence with different
gravitational constant are presented in Fig.3. It is shown that
the effect of the gravitational constant on the momenta of inertia
is obvious. For a canonical neutron star mass ($1.4 M_{\odot}$),
the increment of the momenta of inertia of a star with
$G/G_{0}=0.80$ comparing with that of $G/G_{0}=1$ is about 33$\%$.
If the moment of inertia of neutron stars can be measured
accurately like the mass measurement someday in the future, the
momenta of inertia can provide a probe to investigate the
variation of gravitational constant.

 In summary, we have proposed an effective mechanism to allow
the observed massive neutron star be supported by  soft EOSs, that
is, the decrease of gravitational ``constant" $G$ at super strong
field brings about the mechanism that a soft EOS can support the
astro-observation of massive neutron stars.

\vskip 1cm \noindent

{\large Acknowledgements} \vskip 0.2cm

\noindent This work was supported by the National Natural Science
Foundation of China under Grant No. 10947023  and the Fundamental
Research Funds for the Central Universities (Grant No.
2012ZZ0079). The　project is sponsored　by　SRF　for　ROCS,　SEM.
This research has made use of NASA's Astrophysics Data System.

\end{document}